# Community Detection and Growth Potential Prediction Using the Stochastic Block Model and the Long Short-Term Memory from Patent Citation Networks

Kensei Nakai[1], Hirofumi Nonaka[1], Asahi Hentona[1], Yuki Kanai[1], Takeshi Sakumoto[1],
Shotaro Kataoka[1], Elisa Claire Alemán Carreón[1], Toru Hiraoka[2]

[1]Department of Information and Management Systems Engineering, Nagaoka University of Technology, Nagaoka, Japan
(s163405@stn.nagaokaut.ac.jp, nonaka@kjs.nagaokaut.ac.jp, s173348@stn.nagaokaut.ac.jp, s183348@stn.nagaokaut.ac.jp, s183353@stn.nagaokaut.ac.jp, s183347@stn.nagaokaut.ac.jp, s153400@stn.nagaokaut.ac.jp)
[2]Department of Information Systems, University of Nagasaki, Nagasaki, Japan
(hiraoka@sun.ac.jp)

*Abstract* - **Scoring patent documents is very useful for technology management. However, conventional methods are based on static models and, thus, do not reflect the growth potential of the technology cluster of the patent. Because even if the cluster of a patent has no hope of growing, we recognize the patent is important if PageRank or other ranking score is high. Therefore, there arises a necessity of developing citation network clustering and prediction of future citations. In our research, clustering of patent citation networks by Stochastic Block Model was done with the aim of enabling corporate managers and investors to evaluate the scale and life cycle of technology. As a result, we confirmed nested SBM is appropriate for graph clustering of patent citation networks. Also, a high MAPE value was obtained and the direction accuracy achieved a value greater than 50% when predicting growth potential for each cluster by using LSTM.**

## I. INTRODUCTION

Patents can protect the product innovation of firms from rivals and create a financial benefit [1]. Therefore, patent information is widely used for many studies of technology management such as analysis of technology driven industry [2-3] and real option theory [4]. Patent data is also useful to identify technology trends which help managers plan strategies of R&D, M&A, and other business process. Some researchers applied text-mining techniques to capture technology trends [5-8]. For examples, Nonaka et al, [8] proposed a method for extracting the effect and the technological information from patents by using entropy-based boot-strap and grammar patterns.

While information extraction of patent documents is important to analyze technology, the scoring of patents is also useful for technology management analysis. There are many aspects of patent scoring such as patent application number based index [9]. Among them, the number of cited patents is especially used for quantification of patent value because it shows a degree of attention from other competitors in the same technology field. Hall [10] found that patent citation significantly affects market value. Chang [11] study employs a panel threshold regression model to test whether the patent i-index has a threshold effect on the relationship between patent citations and market value in the pharmaceutical industry. However, simple citation counting, or its expansion, should not be considered the inherent value of importance of different patents. Therefore, it is necessary to assign importance ranks to each patent. To address the issue, a link evaluation method such as PageRank [12] or HITS [13] has been used for calculating importance of patents. Lukach, et al. [14] have proposed computing importance by the PageRank score of patents. According to their study, PageRank patent importance weights differs from the weights based on the number of backward or forward citations. Nonaka, et al. [15] calculated a correlation between the patent score based on HITS, which is similar with Pagerank score, and stock data. They indicated high correlation between the patent score of B to B companies and their time series of stock price value. Additionally, Bruck [16] showed that using the PageRank score of patent documents gives the possibility to make predictions about future technological development. While link evaluation methods are useful for patent scoring, these methods are based on static models and thus do not reflect the growth potential of the technology cluster of the patent. Because, even if the cluster of a patent has no hope of growing, we recognize the patent is important if PageRank or other ranking score is high. Therefore, there arises a necessity of developing citation network clustering and prediction of future citations. In this paper, we proposed a community detection method using the Stochastic Block Model (SBM) [17] and an evaluation of growth potential method using the Long Short-Term Memory (LSTM) [18] deep learning model.
SBM is a generative model for random graphs to group similar nodes to communities. In this research, we adopted a Nested SBM [19] which is suitable for large networks into citation networks. Then we predicted the future citation number of each technology communities using the Long Short Time Memory deep learning model which is widely used for time series data such as financial data [20].

The remainder of this paper is organized as follows. In section 2, we describe the method which is combined between nested SBM as community detection from patent citation network, and LSTM as patent communities' growth analysis. Our experiments on system performance and the result of the analysis are explained in section 3. Section 4 conducts a discussion and evaluation of this system based on the experiment results. Finally, section 5 describes the conclusion of the paper and our future work.

## II. METHODOLOGY

*A. Overview of Our method*

Our proposed method is shown in figure 1.

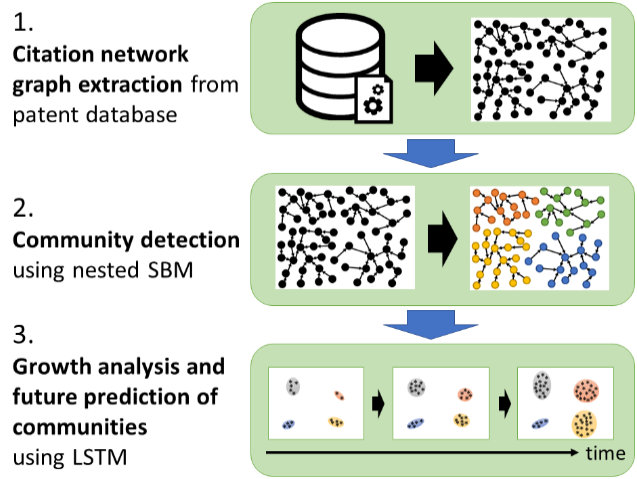

**Figure 1 Outline of our method**

*B. Citation network graph extraction from patent database*

Using the patent information database as an input, we created a graph file of patent citation network in patent field to be analyzed. First, we prepared a patent information database. In this study, we used a Japanese patent database provided by the Institute of Intellectual Property of Japan (IIP) [21]. It contains about 6,000,000 patents from 1960 to 2013 and their examiner citations. After that, it extracts the patent citations network in the patent field to be analyzed from the database and generates a graph file (DOT file). In the graph file, the nodes of the network (patent application number) and the link (quotation relation of the patent) are recorded. This generated graph file is used for input of nested SBM to be performed later.

*C. Community detection using nested SBM*

In this paper, we use Nested SBM [19] to detect patent community. The original SBM classifies nodes into blocks (communities) with edges between each nodes of blocks. The overview of SBM is shown in Figure 2. Nested SBM is a hierarchy SBM model and can detect the best partition

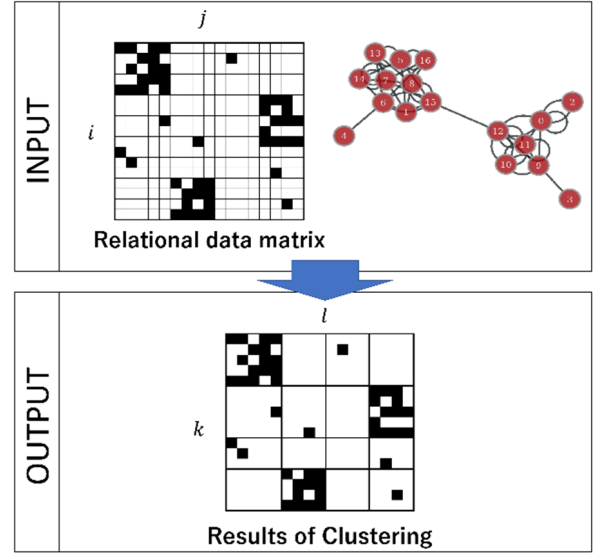

**Figure 2 Overview of SBM**

number automatically. For more details, please refer to reference [19].

*D. Growth analysis and prediction of communities using LSTM*

In this research, we used long short-term memory (LSTM) networks, one of the most advanced deep learning architectures. This model is used in sequence learning tasks such as character recognition, or time series prediction [22-23] to predict citation number of each community. For more details, please refer to reference [20].

## III. RESULTS

*A. Nested-SBM*

In this section, we applied graph clustering algorithm to real patent networks. We analyzed the citation network of the game technology field including "Video games and related technology fields" (IPC code: A63F13) and "Roulette-like ball games and related technology fields" (IPC code: A63F5) for its convenience for understanding the technology and evaluating results. Table I shows the details of the data. In order to conduct the Nested SBM algorithm, we used Python library *graph-tool[1]* developed by Tiago de Paula Peixoto, and clustered patent citation networks.

Figure 3 shows the clustering result by "nested SBM".

---

[1] Graph-tool https://graph-tool.skewed.de/

Table I
Network to Be Analyzed

| Patent classification name | Number of patent nodes | Number of cited links |
|---|---|---|
| Video games | 15631 | 26826 |
| Roulette-like ball games | 29621 | 74529 |

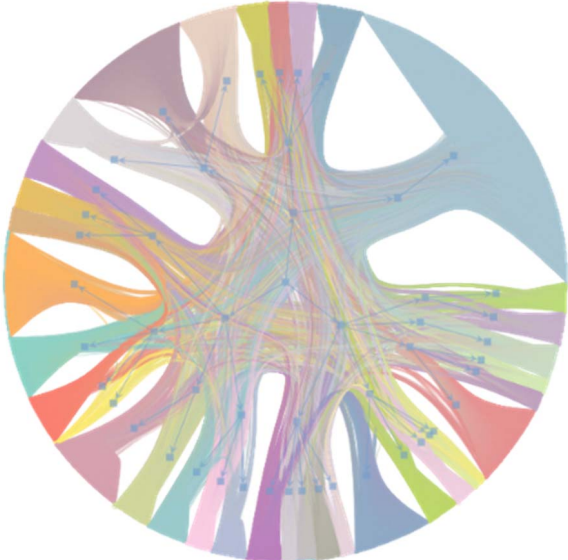

(a) "Video games" (Max. clusters: 36)

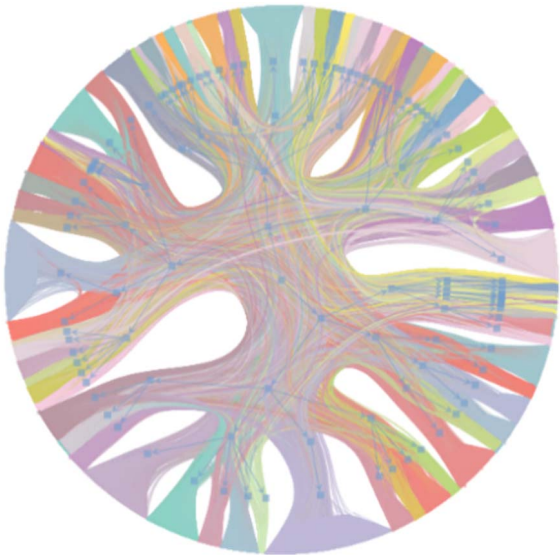

(b) "Roulette-like ball games" (Max. clusters: 88)

**Figure 3 Result of Clustering (Expressed as graphical)**

This figure is a graphical representation of the result of the hierarchical clustering. The cluster is integrated inside the figure as the bottom layer clusters (number of clusters K = target patent number) on the outside of the figure, and it finally shows that it connects to one cluster.

---

[2] TensorFlow https://www.tensorflow.org/

*B. LSTM*

We predicted the number of citations of each community to evaluate the growth potential of the community by LSTM.

We used Python library *TensorFlow*[2]. Our LSTM Neural Network is composed of one input layer, four LSTM layers with memory blocks, and one output layer. We computed the Mean Absolute Percentage Error (MAPE) and accuracy of direction, to measure the effectiveness of different travel speed algorithms.

Table II shows the system performance of our LSTM, and Figure 4 and figure 5 show prediction results of patent citations using LSTM.

The horizontal axis of each graph is the patent application year (the origin of the target data is 1960 as the origin of the horizontal axis), and the vertical axis is the number of citations of the patent.

The red line of each graph is true value, the green line is prediction by training data, and the blue line is prediction by test data.

Table II
System Performance

|  | MAPE (%) | Accuracy of direction (%) |
|---|---|---|
| Video games | 62.06 | 53.94 |
| Roulette-like ball games | 70.82 | 57.1 |

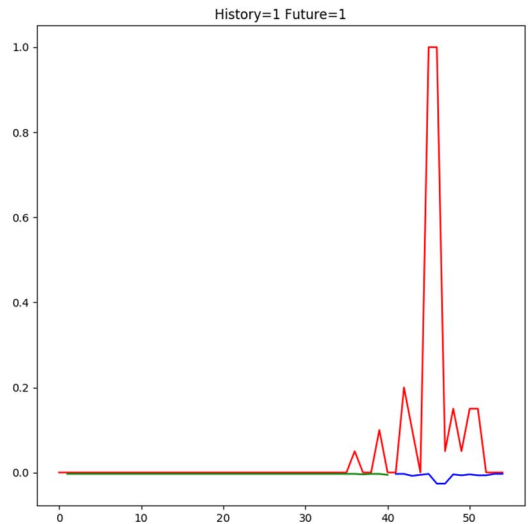

**Figure 4 Result of Prediction (bad example)**

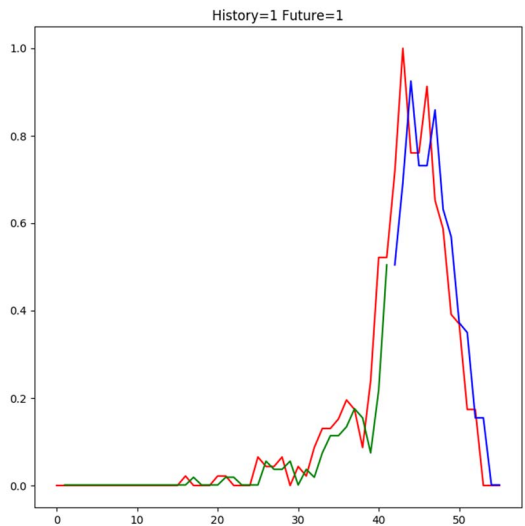

Figure 5  Result of Prediction (good example)

## IV. DISCUSSION

*A. Nested SBM*

Using the results of Nested SBM, we examined the technical field by randomly referring to patent documents of patents belonging to each cluster. As a result, we could find common technical points. Regarding the technology classification "Video Game and related technology fields", 534 patents were applicable to a certain cluster. This cluster is a set of patents of the technique of the game user interface and its technology. The central patent (JP 2003-340042) is a patent concerning "a *pachinko* machine having a function to convey a special event state in the game in an easy-to-understand manner". This patent affects other patents on the way a user interacts with the game. In addition, this cluster includes patents of "screen drawing optimization" and "home game controller".

In another cluster, 340 patents were applicable. The cluster is a collection of patents related to input and output of game machines and control of peripheral devices. The central patent is a patent on a music game system and a play screen structure (JP 平8-305356). This patent is frequently cited in patents of game systems using this technology. In addition to this cluster, there are patents related to "game event occurrence control according to level" and "device control in competitive game".

Regarding another patent classification "ball game like roulette", 897 patents were applicable to a certain cluster. The cluster is a collection of patents related to prevention of tampering with *pachinko* / *pachislot* machines and their peripheral technology. The central patent is a circuit board box (JP 平6-269539) which makes it difficult to tamper with the slot machine. This patent is quoted for slot machines equipped with the circuit board box and patents applying the technology. Besides this, as a patent of this cluster, there is a structure for facilitating the inspection of pachinko machines, and a fraud detection mechanism for pachinko machines. In addition, 968 patents were applicable to another cluster. The cluster is a set of patents related to the production technique of the game machine. The core patent is a slot display technique (P2004-357878). This patent is cited in a technical patent for realizing this method. In addition to this, the cluster has the structure of slot machine parts and the pachinko performance program. In the above examination, we confirmed nested SBM is appropriate for graph clustering of patent citation networks.

*B. Long Short-Term Memory*

As a result of prediction by LSTM, the MAPE value was not good. In particular, it was revealed that the MAPE value was bad in a community that rapidly declined after the patent citation rapidly increased. On the other hand, we can predict patent citations well for communities with long-term trends. According to Table II, prediction of direction showed more than 0.5 accuracy in any technical classification, but there is still a room for improvement.
In the future, it is necessary to consider the number of nodes and the number of layers of LSTM for improvement.

## V. CONCLUSION

In our research, clustering of patent citation networks by Stochastic Block Model was done with the aim of enabling corporate managers and investors to evaluate the scale and life cycle of technology.

As a result, we confirmed nested SBM is appropriate for graph clustering of patent citation networks and the not good MAPE value and direction accuracy more than 0.5 of predicting growth potential of each cluster by using LSTM.

In future work, we improve LSTM to predict growth potential of each cluster. Furthermore, we combined our proposed method and a link evaluation method for investors and corporate managers to predict corporate finance data.